\documentclass[twocolumn,preprintnumbers,amsmath,amssymb,superscriptaddress]{revtex4}
\usepackage{graphicx}
\usepackage{dcolumn}
\usepackage{bm}
\usepackage{soul}
\usepackage{color}
\usepackage{epstopdf}
\usepackage[version=3]{mhchem}
\usepackage{lipsum}
\usepackage[outercaption]{sidecap}
\usepackage{floatrow}
\usepackage{hyperref}

\begin{document}


\title{Photo-to-heat conversion of broadband metamaterial absorbers based on TiN nanoparticles under laser and solar illumination} 
\author{Do T. Nga}
\affiliation{Institute of Physics, Vietnam Academy of Science and Technology, 10 Dao Tan, Ba Dinh, Hanoi 12116, Vietnam}
\author{Anh D. Phan}
\email{anh.phanduc@phenikaa-uni.edu.vn}
\affiliation{Faculty of Materials Science and Engineering, Phenikaa University, Hanoi 12116, Vietnam}
\affiliation{Phenikaa Institute for Advanced Study, Phenikaa University, Hanoi 12116, Vietnam}
\author{Thudsaphungthong Julie}
\affiliation{Faculty of Physics, TNU - Thai Nguyen University of Education, Thai Nguyen, 24000, Vietnam}
\author{Nam B. Le}
\affiliation{School of Engineering Physics, Hanoi University of Science and Technology, 1 Dai Co Viet, Hanoi 10000, Vietnam}
\author{Chu Viet Ha}
\affiliation{Faculty of Physics, TNU - Thai Nguyen University of Education, Thai Nguyen, 24000, Vietnam}
\date{\today}

\begin{abstract}
We theoretically investigate photothermal heating of ultra-flexible metamaterials, which are obtained by randomly mixing TiN nanoparticles in polydimethylsiloxane (PDMS). Due to the plasmonic properties of TiN nanoparticles, incident light is perfectly absorbed in a broadband range (300-3000 nm) to generate heat within these metamaterials. Under irradiation of an 808 nm near-infrared laser with different intensities, our predicted temperature rises as a function of time agree well with recent experimental data. For a given laser intensity, the temperature rise varies non-monotonically with concentration of TiN nanoparticles because the enhancement of thermal conductivity and absorbed energy as adding plasmonic nanostructures leads to opposite effects on the heating process. When the model is extended to solar heating, photothermal behaviors are qualitatively similar but the temperature increase is less than 13 $K$. Our studies would provide good guidance for future experimental studies on the photo-to-heat conversion of broadband perfect absorbers.
\end{abstract}

\maketitle
\section{Introduction}
Plasmonic nanostructures have been exploited in a wide range of applications including photothermal therapy \cite{9,10}, energy storage \cite{11,12,13}, imaging \cite{14,15}, and sensing \cite{16}. Typical plasmonic materials are noble metals since a large number of free electrons on their surface can be collectively excited by incident light to obtain plasmon resonances. Tuning size, shape, interparticle separation distance, and environment affects the local density of surface free-electrons and changes plasmonic properties \cite{19,20}. Although noble metals have good activity and high durability, they are expensive and unstable at high temperatures. Thus, mass production of plasmonic devices remains limited and challenged. Recently, oxides and nitrides have emerged as alternative plasmonic materials in a wide optical range to replace conventional plasmonic materials \cite{17,18}. 

Among transition metal nitrides, TiN has received much attention since its optical properties are equivalent to gold's but TiN is much cheaper and easier to fabricate \cite{17,18}. Although TiN has a smaller carrier density than gold or silver, absorption and extinction spectra of TiN nanostructures are similar to those of gold counterparts in the near infrared and visible regime \cite{21,22,23}. By changing the processing conditions, the dielectric function and other optical properties of TiN are changed \cite{24}. In addition, the melting temperature of TiN is 2930 \ce{^0C}, which is higher than that of gold at 1064 \ce{^0C} and chromium at 1907 \ce{^0C}. Thus, thermodynamic properties of TiN are more stable to operate at high temperatures. 

Compared to other high melting point materials such as tungsten and chromium, TiN has several advantages for designing electronic devices and broadband absorbers. First, TiN is more affordable than W or Cr. Thus, it is more cost-effective to use TiN for fabricating devices in mass production. Second, the deposition process of TiN is also more accessible and can be achieved through physical and chemical vapor deposition techniques such as sputtering or pulsed laser deposition. In contrast, the deposition of W and Cr can be complex and challenging due to high temperatures and corrosive gases \cite{35}. Third, TiN is a more environmentally friendly material than W or Cr, which may cause serious environmental pollution and health effects \cite{31,32,33}. Finally, TiN is compatible with many types of substrate materials \cite{34}. Thus, it is easier to integrate into a variety of device structures.

Recently, TiN and Au nanoparticles have been randomly dispersed in polymer \cite{1} and nanofiber \cite{25}, respectively, to design broadband perfect absorbers having sufficient flexibility. The engineered materials are called ultraflexible metamaterials and used for solar energy harvesting, biomedical applications, photo-to-thermal conversion, and energy storage. The flexibility and stretchability of these metamaterials create more multifunctional applications than their rigid counterparts. Such applications require deep understandings of physical mechanisms in the photothermal process to effectively optimize performance. Although one can employ finite element simulations to understand spatial and temporal temperature variation \cite{1} in the ultraflexible metamaterials, quantitative comparisons have not been achieved. The simulation also hardly reveals the time dependence of the light-to-heat conversion. Thus, it is necessary to exploit theoretical approaches and validate their limitations. 

In a recent work \cite{41}, we proposed a new model to understand how to control nanoparticle self-assembly and the spatial temperature distribution of ultraflexible metamaterials composed of aramid nanofibers and Au nanoparticles under laser illumination. We indicated that the structural configuration of nanoparticle assembly is thermally affected but plasmonic coupling can be ignored. Theoretical steady-state temperature distribution and the laser-intensity dependence of the hottest temperature agree quantitatively well with experimental data. However, a simple arithmetic average for calculating the effective thermal conductivity, specific heat capacity, and dielectric function of composites in Ref. \cite{41} is not a good approximation \cite{8,27}. In addition, several unsolved questions are: (i) Can the time-dependent temperature predicted by this model compare with experiments? (ii) Can this model apply to solar energy harvesting? (iii) How can we use this approach to optimize behaviors of the photothermal heating?

This article addresses the above questions and other related problems by applying the proposed model in Ref. \cite{41} with better effective medium approximations for properties and calculations for absorption coefficient to mimic photothermal experiments on TiN-based ultra-flexible metamaterials in Ref. \cite{1}. After calculating the time-dependent thermal gradient when exposed to laser radiation, we compare numerical results with experiments and validate assumptions. Varying concentrations of particles in a wide range suggest us how to optimize the photothermal heating. Then, we extend the model to investigate conversion of optical energy to thermal energy under solar irradiation. Applications of our approach to other systems are also discussed.

\section{Theoretical background}
Motivated by the metamaterials and light-to-heat conversion in Ref.\cite{1}, we consider the photothermal heating of a random mixture of TiN nanoparticles and PDMS. Under laser illumination, TiN nanoparticles absorb the light energy, perfectly dissipate into heat, and increase the temperature of the surrounding medium. Based on a heat energy balance equation, one derives an expression of thermal response of a semi-infinite substrate \cite{6,7} to calculate the temporal and spatial temperature rise. This expression is
\begin{widetext}
\begin{eqnarray}
\Delta T(r,z,t)=\frac{I_0(1-\mathcal{R})\alpha}{2\rho_d c_d}\int_0^t \exp\left(-\frac{\beta^2r^2}{1+4\beta^2\kappa t'} \right)\frac{e^{\alpha^2\kappa t'}}{1+4\beta^2\kappa t'}\left[e^{-\alpha z}\ce{erfc}\left(\frac{2\alpha\kappa t'-z}{2\sqrt{\kappa t'}} \right)+ e^{\alpha z}\ce{erfc}\left(\frac{2\alpha\kappa t'+z}{2\sqrt{\kappa t'}} \right)\right]dt',
\label{eq:3}
\end{eqnarray}
\end{widetext}
where $z$ is the depth direction parallel with the incident field, $r$ is the radial distance in the horizontal plane, $I_0$ is the laser intensity, $\alpha$ is the effective absorption coefficient, $\mathcal{R}\approx 0$ is the reflectivity, and $\beta = 500$ $\ce{m^{-1}}$ is the inverse of the laser spot radius \cite{1}, $\kappa=K_d/\rho_dc_d$ is the thermal diffusivity, $K_d$ is the thermal conductivity, $c_d$ is the specific heat capacity, and $\rho_d$ is the mass density. Since TiN nanoparticles are randomly dispersed in PDMS, the effective thermal conductivity of the composites is estimated using its components and the Hamilton-Crosser model equation \cite{27,28}, which is
\begin{widetext}
\begin{eqnarray}
K_d = K_{PDMS}\frac{K_{TiN}+(n-1)K_{PDMS}+(n-1)\Phi\left(K_{TiN}-K_{PDMS}\right)}{K_{TiN}+(n-1)K_{PDMS}-\Phi\left(K_{TiN}-K_{PDMS}\right)},
\label{eq:4-1}
\end{eqnarray}
\end{widetext}
where $K_{PDMS} = 0.16$ W/m/K \cite{1}, $K_{TiN} = 60$ W/m/K \cite{1}, $n$ is a parameter capturing effects of finite size and shape of the nanoparticles, and $\Phi = N4\pi R^3/3$ is the volume fraction of TiN particles with $N$ and $R$ being the number of particles per volume and the radius of particles, respectively. For spherical nanoparticles, $n$ is set to be 3 \cite{27}. The Hamilton-Crosser model was found to provide good quantitative descriptions for the effective thermal conductivity \cite{27}.

Meanwhile the effective specific heat capacity and mass density are calculated by \cite{27}
\begin{widetext}
\begin{eqnarray}
c_d &=&  c_{PDMS}\left(1-\frac{\rho_{TiN}\Phi}{\rho_{PDMS}(1-\Phi)+\rho_{TiN}\Phi} \right)+c_{TiN}\frac{\rho_{TiN}\Phi}{\rho_{PDMS}(1-\Phi)+\rho_{TiN}\Phi},\nonumber\\
\rho_d &=&  \rho_{PDMS}(1-\Phi) + \rho_{TiN}\Phi,
\label{eq:4}
\end{eqnarray}
\end{widetext}
where $\rho_{PDMS}=970$ kg/\ce{m^3}, $\rho_{TiN}=5400$ kg/\ce{m^3}, $c_{PDMS}=1460$ \ce{J/kg/K}, $c_{TiN}=533$ \ce{J/kg/K}\cite{1}.

There are two main methods to determine the effective absorption coefficient. First, one can use the Maxwell-Garnett approximation \cite{8}, which gives
\begin{widetext}
\begin{eqnarray}
\alpha=\frac{4\pi}{\lambda}\ce{Im}\left(\sqrt{\varepsilon_{PDMS}\frac{\varepsilon_{TiN}\left(1+2\Phi\right)+2\varepsilon_{PDMS}\left(1-\Phi\right)}{\varepsilon_{TiN}\left(1-\Phi\right)+\varepsilon_{PDMS}\left(2+\Phi\right)}}\right),
\label{eq:5}
\end{eqnarray}
\end{widetext}
where $\varepsilon_{PDMS}\approx1.96$ \cite{1} and $\varepsilon_{TiN}$ are the dielectric function of PDMS and TiN, respectively, and $\lambda$ is the incident wavelength. The effective dielectric function of the composite system provided by the Maxwell-Garnett model is known to be better than simply averaging dielectric functions of components in the composite \cite{8}. The dielectric function of TiN can be described using a generalized Drude-Lorentz model over a wide range of frequency \cite{2}
\begin{eqnarray}
\varepsilon_{TiN}(\omega)=\varepsilon_{\infty}-\frac{\omega_{p}^2}{\omega(\omega+i\Gamma_D)}+\sum_{j=1}^2\frac{\omega_{L,j}^2}{\omega_{0,j}^2-\omega^2-i\gamma_j\omega},
\label{eq:6}
\end{eqnarray}
where $\omega_{p} \approx 7.38$ eV is the plasma frequency, $\varepsilon_{\infty}=5.18$ is the infinite frequency permittivity, $\Gamma_D \approx 0.26$ eV is the Drude damping parameter, $\omega_{L,1}\approx 6.5$ \ce{eV} and $\omega_{L,2}=1.5033$ \ce{eV} are the Lorentz oscillator strengths, $\omega_{0,1}=4.07$ \ce{eV} and $\omega_{0,2}=2.02$ \ce{eV} are the Lorentz energies, $\gamma_1 = $ 1.42 \ce{eV} and $\gamma_1 = $ 1.42 \ce{eV} are the Lorentz damping parameters \cite{2}. Second, according to the Beer–Lambert law, the effective absorption coefficient is
\begin{eqnarray}
\alpha=NQ_{ext}=\frac{3\Phi Q_{ext}}{4\pi R^3},
\label{eq:7}
\end{eqnarray}
where $Q_{ext}$ is the extinction cross section of a TiN nanoparticle. Two main mechanisms contributing to the extinction are absorption and scattering. The validity of Eqs. (\ref{eq:5}) and (\ref{eq:7}) is discussed in the following section.

\section{Results and discussions}
\begin{figure}[htp]
\includegraphics[width=9cm]{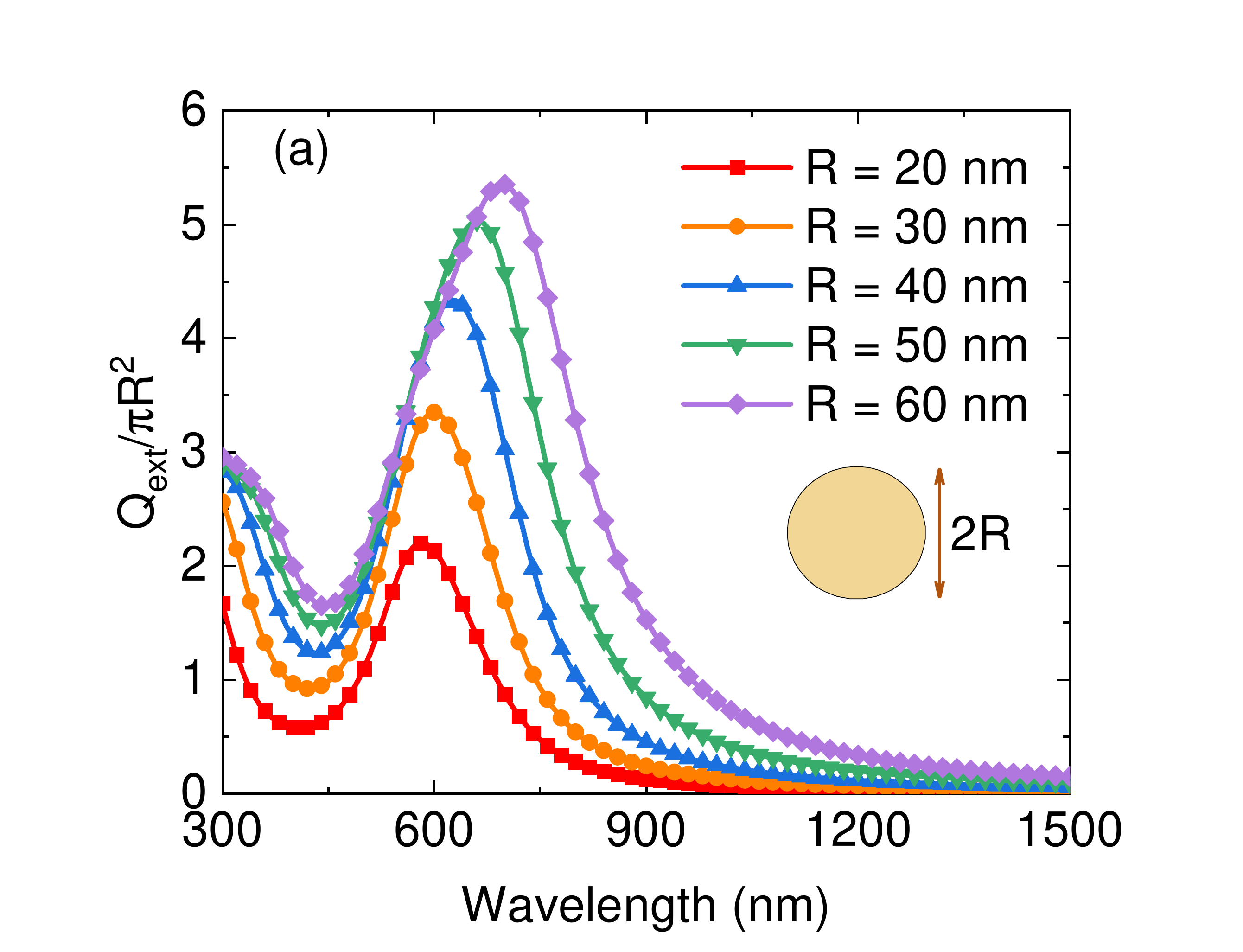}
\includegraphics[width=9cm]{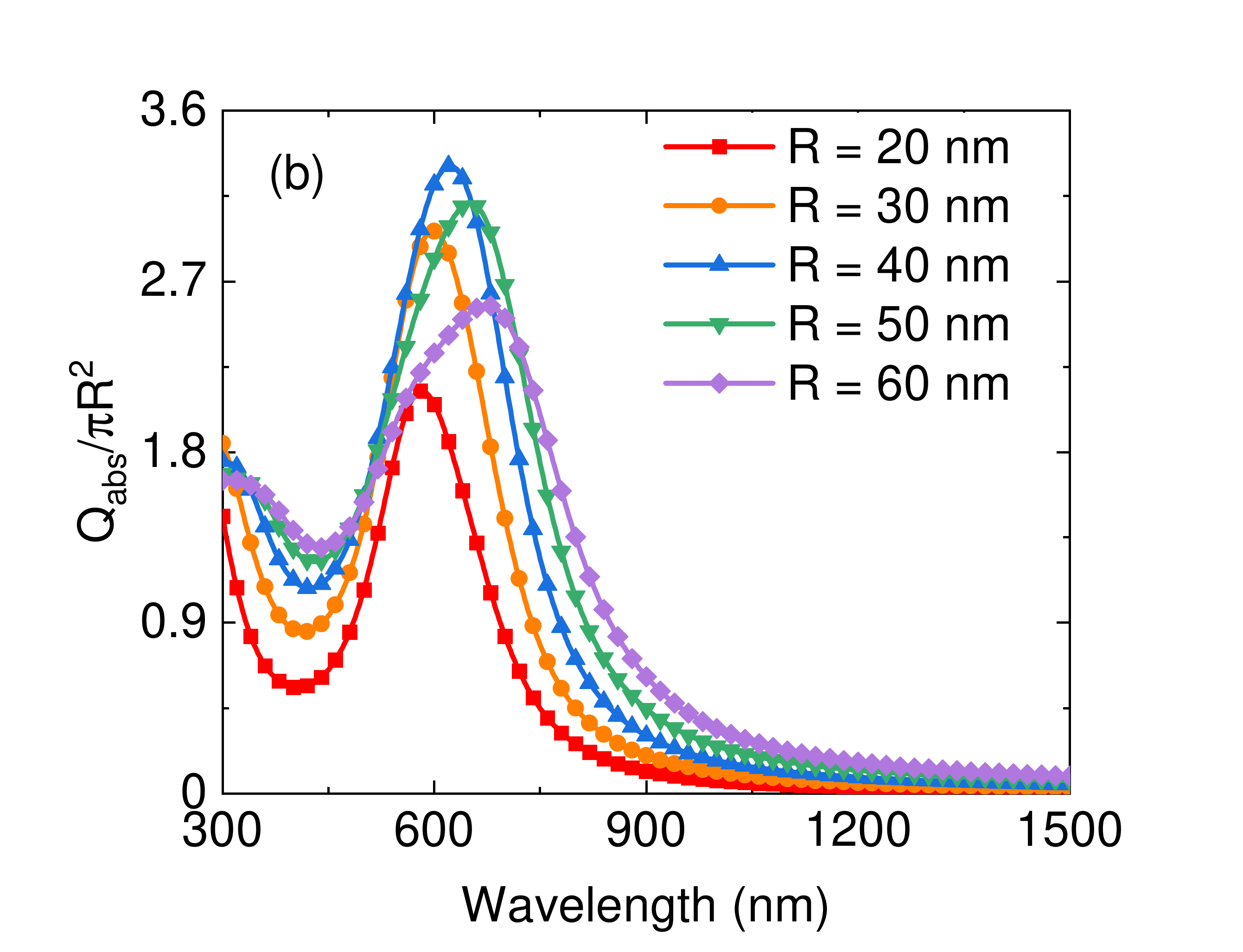}
\caption{\label{fig:1}(Color online) The wavelength dependence of (a) the extinction cross section and (b) absorption cross section of TiN nanoparticles normalized by its area at different radii.}
\end{figure}

\begin{figure}[htp]
\includegraphics[width=9cm]{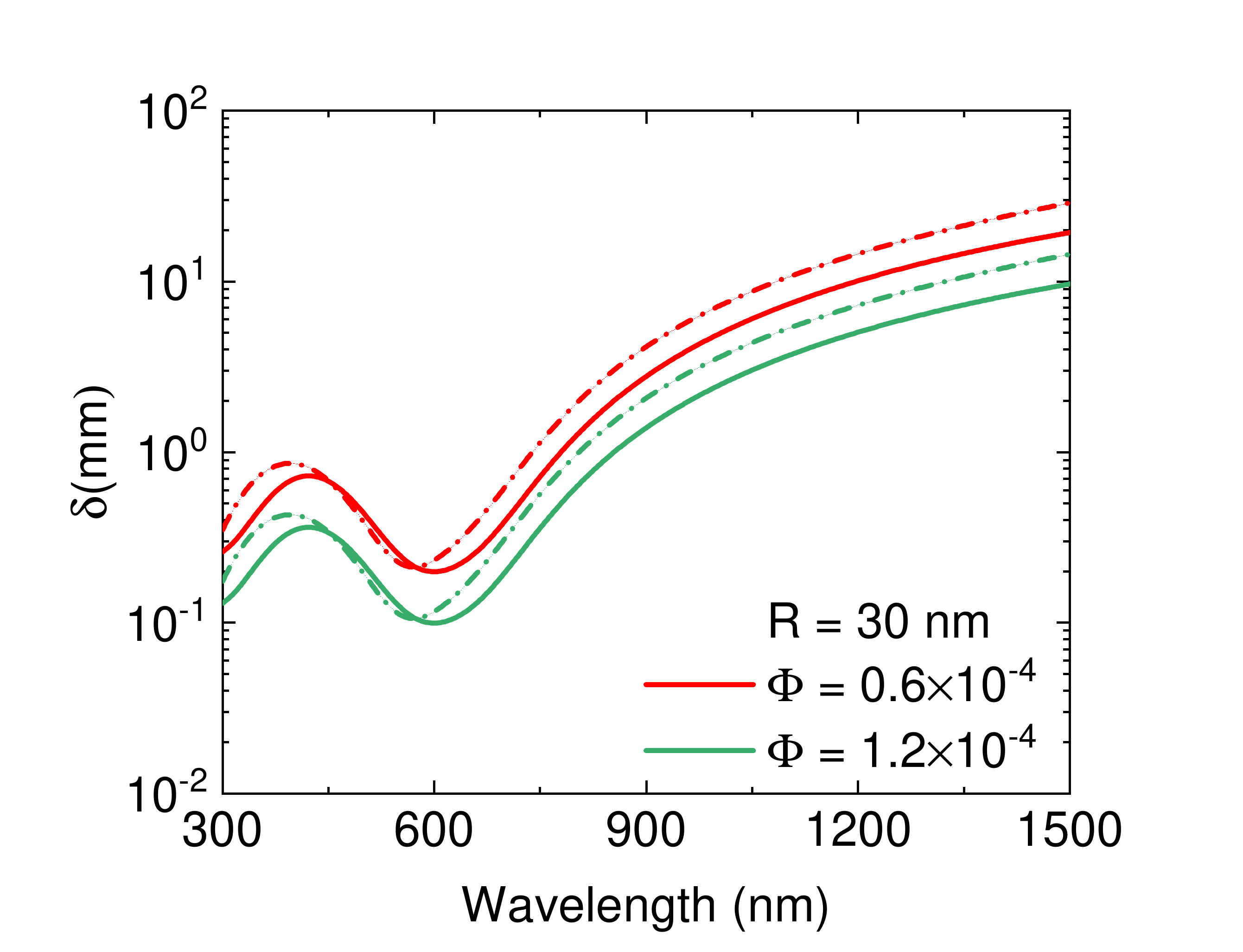}
\caption{\label{fig:2}(Color online) The absorption length as a function of wavelength calculated using the Beer–Lambert law (solid curves) and Maxwell-Garnett approximation (dashed-dotted curves) at different volume fractions of 30-nm TiN nanoparticles.}
\end{figure}

Before investigating the temperature increase, we theoretically study $Q_{ext}$ to understand variations of the absorption coefficient. We employ Mie theory \cite{6,8} to calculate the normalized extinction cross section at different sizes of TiN nanoparticles and present numerical results in Fig. \ref{fig:1}a. The maximum value of the normalized extinction spectrum approximately raises from 2.2 to 5.4 when $R$ increases from 20 nm to 60 nm. Increasing the particle size also red-shifts the spectrum. This variation behaves in the same manner as the normalized absorption ($Q_{abs}/\pi R^2$) as shown in Fig. \ref{fig:1}b. Our numerical results clearly indicate that the extinction cross section is mainly contributed by the absorption for small nanoparticles. The scattering contribution to the extinction becomes dominant as the particle size increases. A reduction of the extinction cross section at long wavelengths allows large transmission of light through the system and, thus, the absorption length, $\delta=1/\alpha$, becomes longer. At a certain wavelength, adding TiN nanoparticles to the composite enhances the absorption and reduces $\delta$. The analysis is consistent with calculations in Fig. \ref{fig:2}a. 

The absorption lengths as a function of wavelength of flexible metamaterials of 30-nm TiN nanoparticles determined by both the effective medium approximation and the Beer–Lambert law are shown in Fig. \ref{fig:2}. These two theoretical approaches provide numerical results close to each other for a given volume fraction. One can capture more finite-size effects in Eq. (\ref{eq:6}) by modifying the Drude damping parameter $\Gamma_D \rightarrow \Gamma_D+Av_F/R$, where $v_F$ is the Fermi velocity and $A$ is an adjustable parameter describing the change in the mean free path of electrons \cite{29,30}. However, our goal is to develop a minimalist approach to reasonably predict the absorbed energy without any adjustable parameter. Thus, we use Eq. (\ref{eq:6}) without any modification. The Maxwell-Garnett approximation can be analytically derived by considering contributions of the transverse electric and magnetic dipole mode to the extinction cross-section. Meanwhile, the full Mie calculations take into account multipole contributions of the surface plasmon to the absorption and scattering. It means that finite-size effects of nanostructures are not fully encoded in the Maxwell-Garnett approximation. For this reason, we use the Beer–Lambert law to calculate the effective absorption coefficient and other calculations.

\begin{figure}[htp]
\includegraphics[width=9cm]{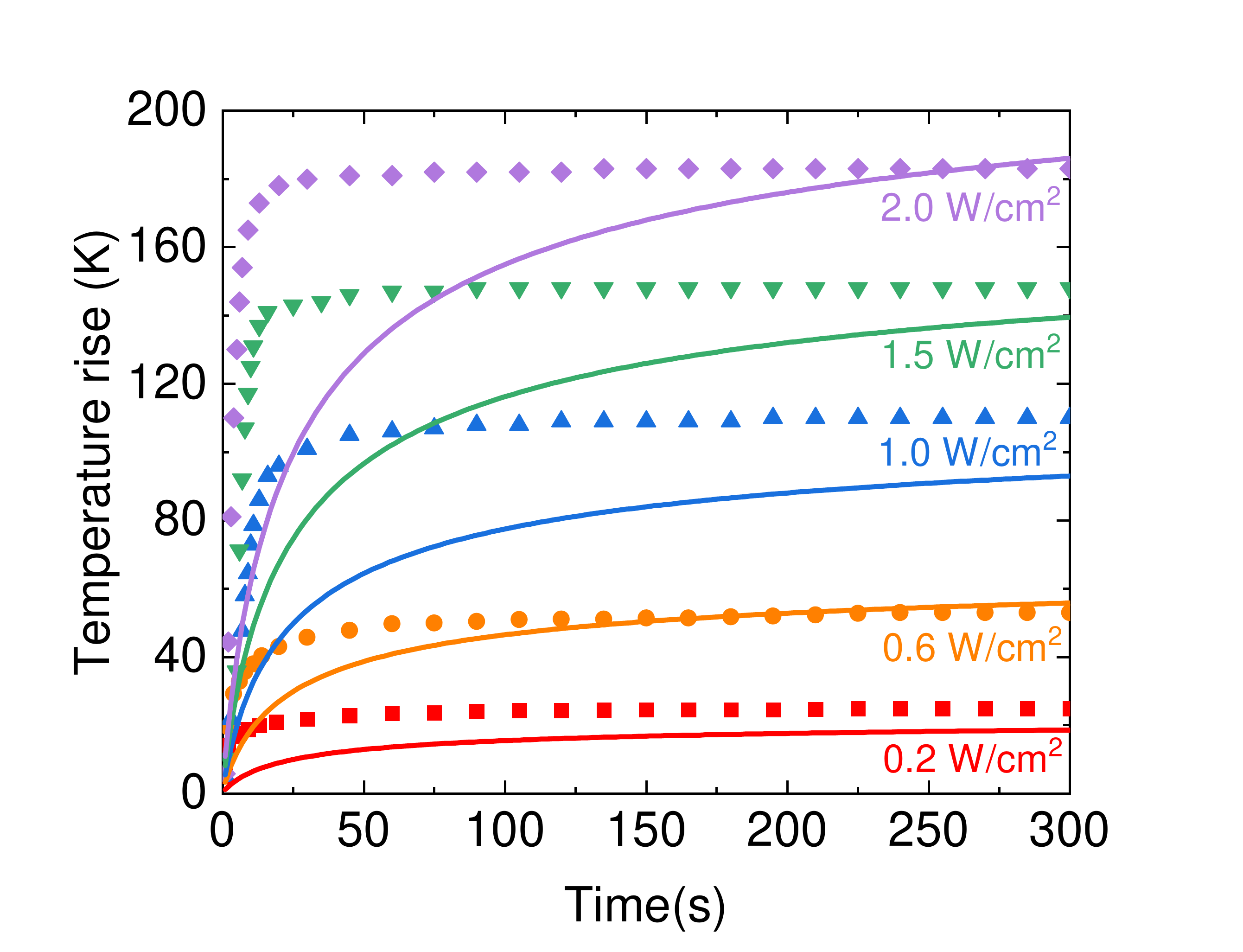}
\caption{\label{fig:3}(Color online) Time dependence of the largest temperature rise at the surface ($r=z=0$) of ultraflexible metamaterials with $\Phi=0.8\times 10^{-4}$ of 30-nm TiN nanoparticles under 808-nm laser irradiation at various laser intensities. Data points and solid curves correspond to experimental data \cite{1} and our numerical results, respectively.}
\end{figure}

Under laser illumination, the photo-to-heat conversion of TiN nanoparticles increases temperature of ambient medium. Figure \ref{fig:3} shows the time-dependent temperature rise at the surface with different intensity laser radiation calculated using Eqs. (\ref{eq:3}), (\ref{eq:4-1}), (\ref{eq:4}), and (\ref{eq:7}). To compare with the metamaterials fabricated in Ref.\cite{1}, the radius is hereafter fixed at $R = 30$ nm. Theoretical predictions and experimental data in Ref. \cite{1} corresponding to $I_0=0.2$ and 0.6 \ce{W/cm^2} are relatively close to each other. Higher intensity laser radiation causes a larger growth of the surface temperature. For $I_0 \geq 1$ \ce{W/cm^2} and $t\geq25s$, $\Delta T(r=z=0,t) > $ 100 $K$ and properties of PDMS polymer can be thermally varied. However, our approach and other simulations \cite{3,4,5} assume that the thermal conductivity, specific heat capacity, and mass density remain unchanged as increasing temperature. This explains why theory predictions deviate from experimental counterparts. In addition, equation (\ref{eq:3}) indicates a linear correlation between the temperature rise $\Delta T$ and the laser intensity $I_0$. This variation closely agrees with photothermal results in previous works \cite{25,36,37}.

For known materials, this approach provides a predictive model, which can well describe photothermal experiments in a minimalist manner. If the thermal conductivity of plasmonic nanoparticles is unknown, one can tune this parameter to obtain quantitative agreement between theory and experiments at low laser intensities and, thus, determine the thermal conductivity of nanostructures.

\begin{figure}[htp]
\includegraphics[width=9cm]{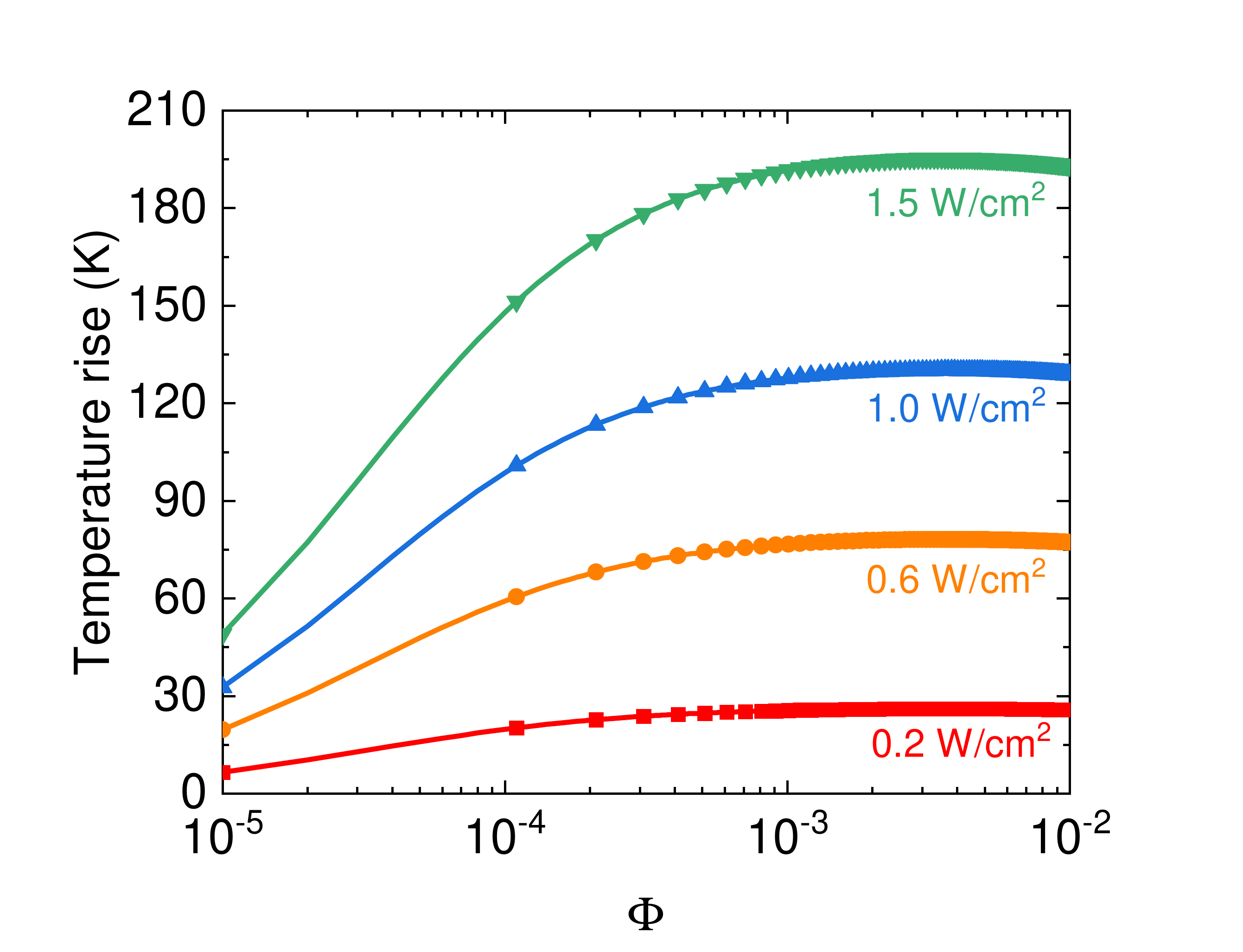}
\caption{\label{fig:4}(Color online) The largest temperature rise at the surface ($r=z=0$) of ultraflexible metamaterials at $t=300$ $s$ as a function of volume fraction of 30-nm TiN nanoparticles under 808-nm laser irradiation at various laser intensities.}
\end{figure}

The concentration of plasmonic nanoparticles has a significant and nontrivial effect on the photothermal heating. From pristine PDMS, adding TiN nanoparticles to the polymer enhances the absorbed energy. Thus, for a fixed laser intensity, $\Delta T(r=z=0,t=300s)$ grows when $\Phi$ increases from 0 to $\sim 3\times10^{-3}$ as shown in Figure \ref{fig:4}. At higher volume fractions ($\Phi\geq 3\times10^{-3}$), although the presence of more plasmonic nanoparticles absorbs more light energy, it enhances the thermal conductivity and reduces the temperature rise. However, the drop of $\Delta T(r=z=0,t=300s)$ at high volume fractions is relatively small. The result suggests that the photothermal heating can be approximately assumed to remain unchanged when the volume fraction is greater than a critical value. In addition, optimizing the photothermal effect requires consideration of the density and size of plasmonic nanoparticles.

\begin{figure}[htp]
\includegraphics[width=9cm]{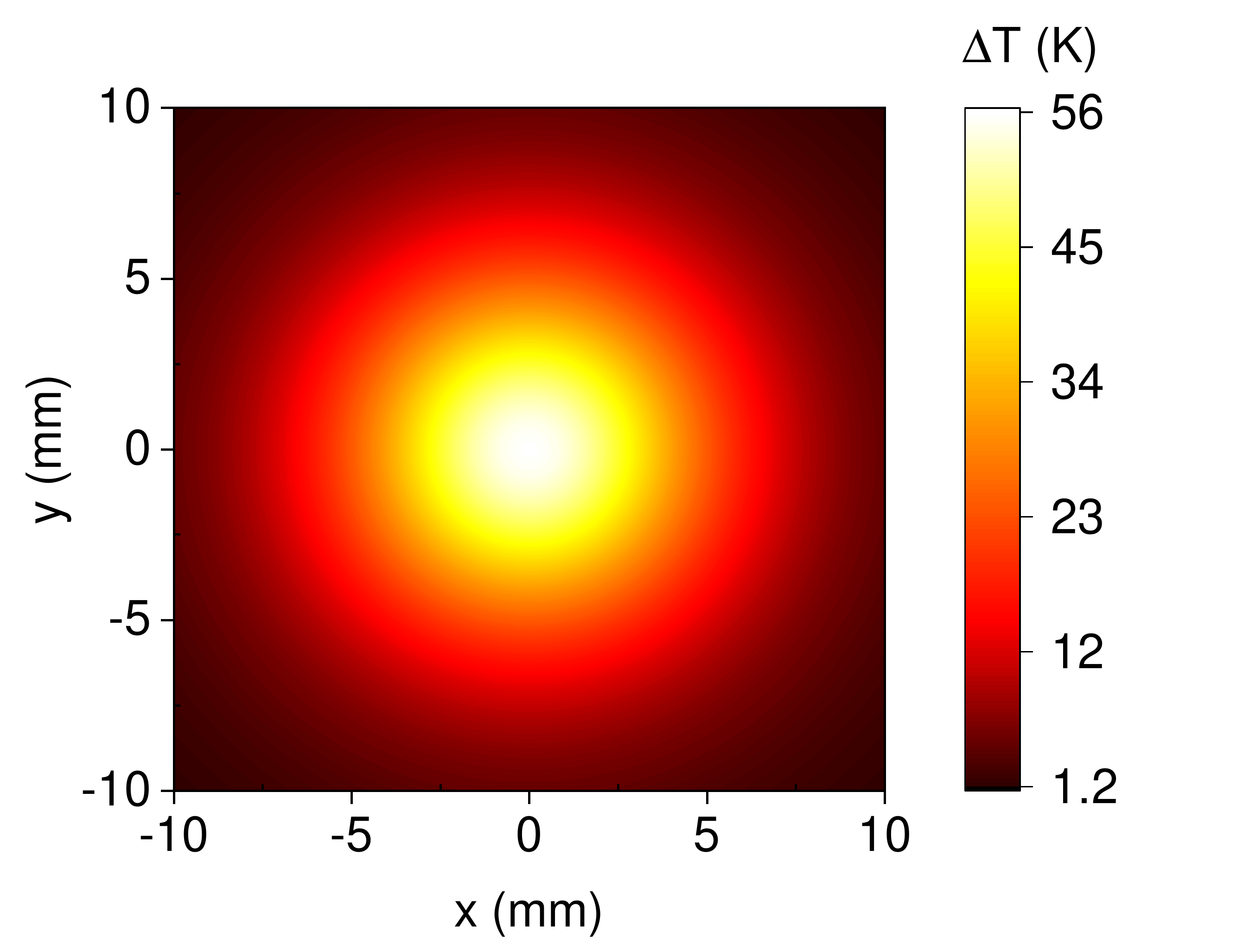}
\caption{\label{fig:5}(Color online) Spatial contour plot of the temperature increase on the surface ($z=0$) of the ultra-flexible metamaterial exposed by a laser spot of 4 mm with the intensity $I = 0.6$ $W/cm^2$. The packing fraction of 30-nm TiN nanoparticle is $\Phi=0.8\times 10^{-4}$.}
\end{figure}

Equation (\ref{eq:3}) also allows us to determine spatial temperature distribution on the surface of the metamaterials irradiated by laser with the intensity $I_0 = 0.6$ $W/cm^2$. As shown in Fig. \ref{fig:5}, the temperature rise at the hottest spot is $\Delta T(r=z=0) = 56$ $K$. The diameter of the hot area defined by $\Delta T(r,z=0) \geq 40$ $K$ is about 6 mm. These calculations can be compared with the thermal gradient measured by infrared camera. One can experimentally test our model in different ways.

When we consider the ultra-flexible metamaterial is exposed under sunlight, Eq. (\ref{eq:3}) can be rewritten as
\begin{widetext}
\begin{eqnarray}
\Delta T(r,z,t)=\int_{\lambda_{min}}^{\lambda_{max}}\frac{E_\lambda\alpha}{2\rho_d c_d}d\lambda\int_0^t \exp\left(-\frac{\beta^2r^2}{1+4\beta^2\kappa t'} \right)\frac{e^{\alpha^2\kappa t'}}{1+4\beta^2\kappa t'}\left[e^{-\alpha z}\ce{erfc}\left(\frac{2\alpha\kappa t'-z}{2\sqrt{\kappa t'}} \right)+ e^{\alpha z}\ce{erfc}\left(\frac{2\alpha\kappa t'+z}{2\sqrt{\kappa t'}} \right)\right]dt',
\label{eq:8}
\end{eqnarray}
\end{widetext}
where $E_\lambda$ is the AM1.5 global solar spectrum, $\lambda_{min}=280$ nm and $\lambda_{max}=3000$ nm are the lower and upper limit of the wavelength range of incident radiation, respectively.

\begin{figure}[htp]
\includegraphics[width=9cm]{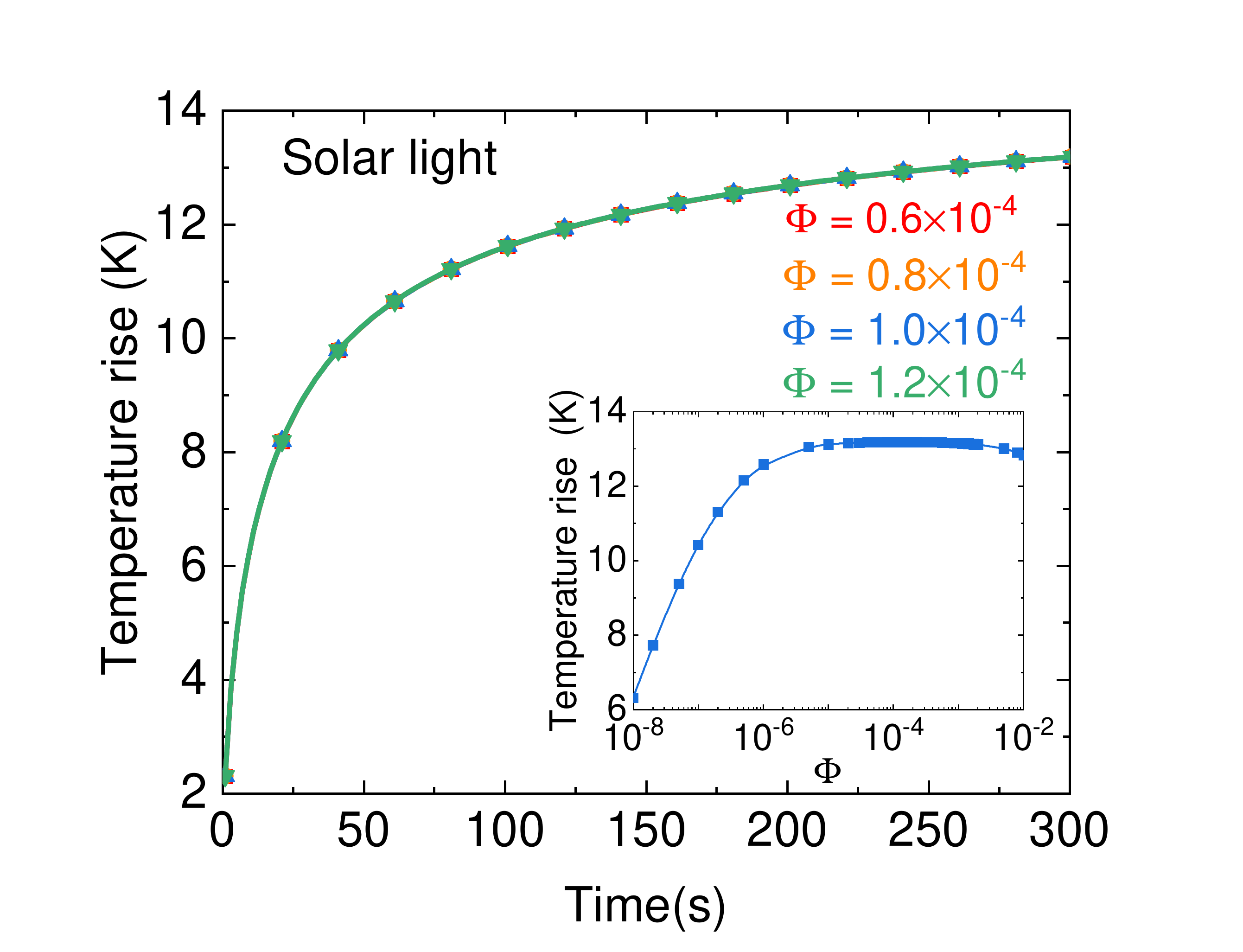}
\caption{\label{fig:6}(Color online) Time dependence of the largest temperature rise on the surface of the ultra-flexible metamaterial irradiated by a solar spectrum with a spot size of 4 mm and different packing fractions of 30-nm TiN nanoparticle. The inset shows the solar-induced temperature rise at $t = 300$ $s$ as a function of the packing fraction.}
\end{figure}

Figure \ref{fig:6} shows the temperature rise at the surface of the metamaterials as a function of time with several values of $\Phi$. The spot size of the solar radiation is still kept at 4 cm. At a given volume fraction, the temperature monotonically increases with time during the irradiation period. When $\Phi$ increases from $0.6\times10^{-4}$ to $1.2\times10^{-4}$, $\Delta T(z=0,t)$ nearly remains unchanged with changing the volume fraction. Physically, the absorbed energy and thermal conductivity still have opposite effects on the temperature rise over a wide range of wavelength. One can expect a non-monotonic variation of $T(z=0,t)$ with $\Phi$ in the same manner as the results in Fig. \ref{fig:4}. We carry out the theoretical calculations and present them in the inset of Fig. \ref{fig:6}. We find that after turning on the solar light for 300 s, the temperature rise non-monotonically varies with $\Phi$ and reaches the maximum at $\Phi\approx10^{-4}$. However, $\Delta T(r=z=0,t=300s)$ slightly changes as $\Phi \geq 10^{-5}$. This result clearly explains why four curves of $\Delta T(r=z=0,t)$ in the mainframe of Fig. \ref{fig:6} overlap as varying the volume fraction. It also suggests that the same photothermal heating is obtained when as $\Phi \geq 10^{-5}$. This temperature rise is much smaller than the case of laser irradiation above. The main reason is the intensity of solar radiation is $\int_{\lambda_{min}}^{\lambda_{max}}E_\lambda d\lambda\approx1000$ \ce{W/m^2}, while the investigated laser intensities are greater than 2000 \ce{W/m^2}.

By tuning the upper and low limits of the integral of Eq. (\ref{eq:8}), we determine the contribution of separated wavelength regimes to $\Delta T(z,t)$. We find that the optical energy of UV range contributes about 0.61 $K$ of the temperature rise. The heating effect is sufficiently small to damage/destroy structures in the ultraflexible metamaterials. Under solar irradiation, only one possibility for degradation of these metamaterials or outdoor devices is the UV-induced bond breaking.

When PDMS in our metamaterials is replaced with water, the system becomes TiN nanoparticles of 30 nm radius randomly dispersed in a water solution. Although the state of matter changes from solid to liquid and the dielectric function of water is frequency dependent, our above approach and equations can be still applied to investigate the photothermal heating in the aqueous solutions under laser and solar illumination. 

For water, the thermal conductivity is 0.6 W/m/K, the mass density is 1000 \ce{kg/m^3}, and the specific heat is 4200 J/kg/K. The absorption coefficient, $\alpha_w$, and the dielectric function of water can be determined using experimental data in Ref. \cite{26}. Thus, the effective absorption coefficient of TiN nanoparticle solution is 
\begin{eqnarray}
\alpha=\alpha_w + NQ_{ext}=\alpha_w + \frac{3\Phi Q_{ext}}{4\pi R^3}.
\label{eq:10}
\end{eqnarray}
Note that although based on the Beer-Lambert law, the explicit difference between Eq. (\ref{eq:10}) and Eq. (\ref{eq:7}) is the effective absorption coefficient of medium $\alpha_w$. For PDMS, the dielectric function is supposed to be a real constant and has no imaginary part. It means that the medium does not absorb the incident light and the effective absorption coefficient of PDMS is zero. Meanwhile, $\alpha_w > 0$ and pure water can absorb light energy.

\begin{figure}[htp]
\includegraphics[width=9cm]{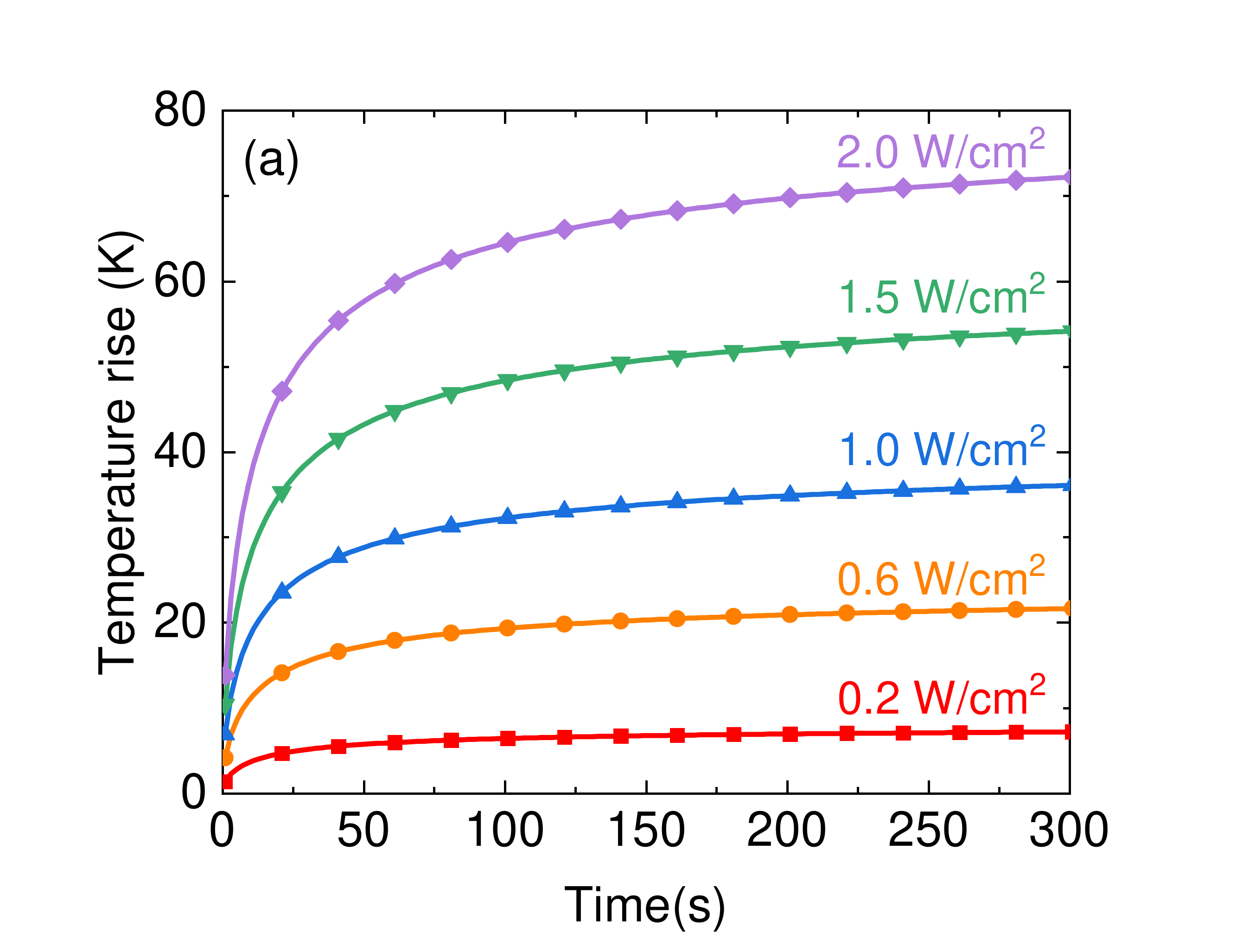}
\includegraphics[width=9cm]{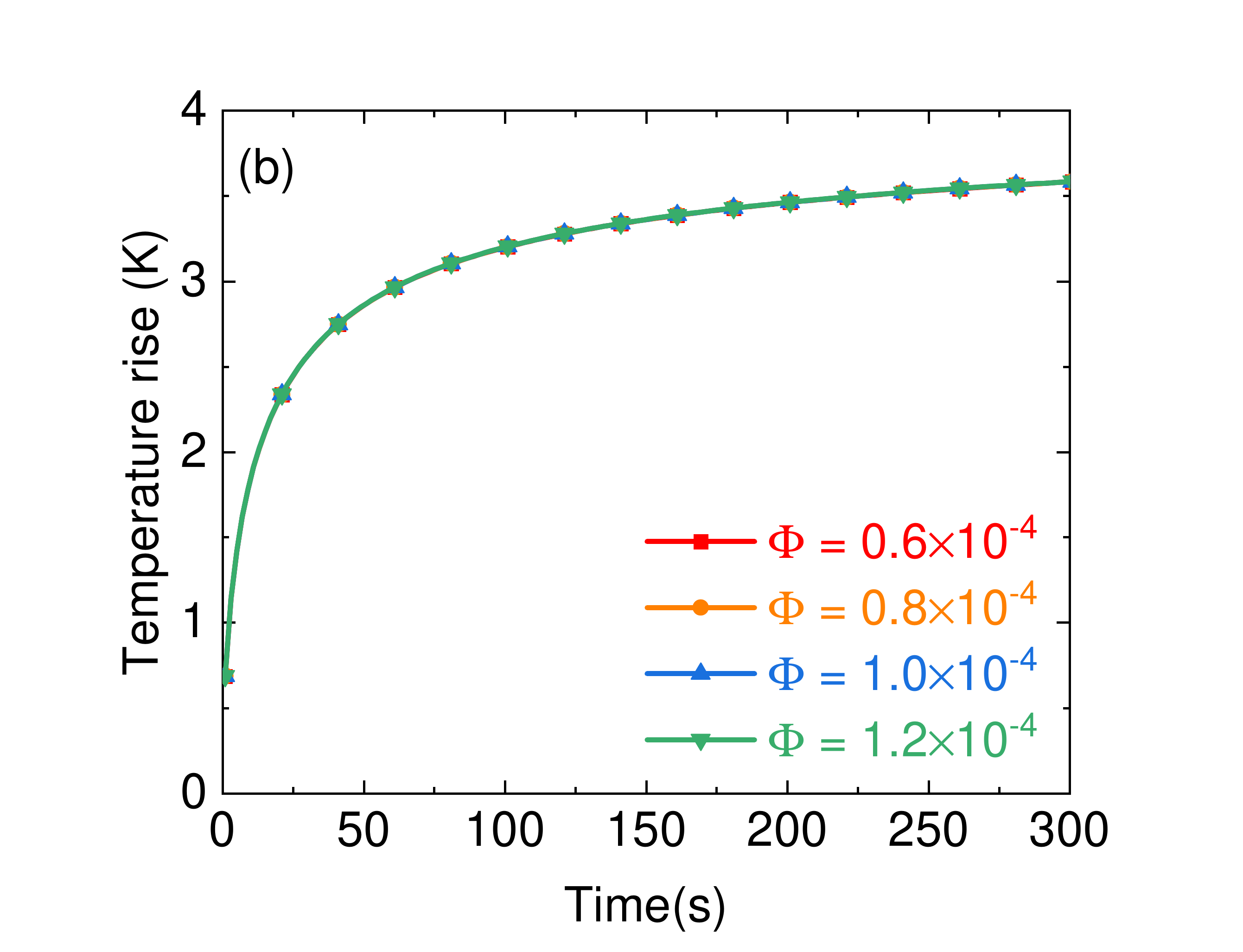}
\caption{\label{fig:7}(Color online) Time dependence of the largest temperature rise on the surface of TiN nanoparticle solutions irradiated by (a) 808-nm laser with different laser intensities and (b) a solar light with a spot size of 4 mm. The average radius of TiN nanoparticles is 30 nm.}
\end{figure}

Figure \ref{fig:7}a shows our numerical calculations for the temperature rise at the hottest spot of TiN nanoparticle solutions having $\Phi=0.8\times 10^{-4}$ under different laser intensities. At a given laser intensity, $\Delta T(r=z=0,t)$ of the TiN nanoparticle solutions is much less than that of our studied metamaterials (results in Fig. \ref{fig:3}). A main reason is that thermal conductivity and specific heat of water are larger than those of PDMS. Water requires more thermal energy to increase temperature compared to PDMS. The same observation can be found in Fig. \ref{fig:7}b when the 808-nm laser light is replaced by the solar light. The temperature rise in the solution is significantly depressed in comparison with the case of ultraflexible metamaterials as presented in Fig. \ref{fig:6}. Again, $\Delta T(r=z=0,t)$ nearly remains unchanged within the range of volume fraction of $0.6\times10^{-4}-1.2\times10^{-4}$. Our model can be applied to other host materials. 

\section{Conclusions}
We have theoretically studied optical properties and the temperature rise of TiN-based ultra-flexible metamaterials under illumination of laser and solar light. Approximations used to compute the effective thermal conductivity, specific heat capacity, and dielectric functions in this work are more accurate than those used in Ref. \cite{41}. Metamaterials with a small TiN concentration absorb a small amount of light and have slight heating. Increasing the TiN concentration enhances both light absorption and thermal conductivity, and leads to a non-monotonic variation of the light-induced temperature rise. As the volume fraction is greater than a critical value, the temperature is slightly varied. The photo-to-heat conversion at a large volume fraction of TiN nanoparticles can be quantitatively equivalent to that at a small volume fraction. We can minimize the concentration of nanostructures dispersed in PDMS but still obtain an appropriate heating performance for photothermal and solar harvesting applications. At low-laser intensities ($I_0\leq 0.6$ \ce{W/cm^2}), our calculations for the time dependence of the surface temperature rise describe well experiments in Ref. \cite{1}. It is the first time the predictions of $\Delta T(t)$ of ultraflexible metamaterials have been compared with experiments. It indicates that the proposed model is reliable. At higher-laser intensities, an increase of temperature greater than 100 $K$ could change the thermal properties of the metamaterials and lead to remarkable deviations between theory predictions and experiments. Since the solar light intensity is 0.1 \ce{W/cm^2}, the temperature rise at the hottest spot of the metamaterials after $t = 300$s of solar light irradiation is less than 13 $K$ but it non-monotonically varies with the volume fraction in the same manner as the case of laser irradiation. This finding suggests how to optimize the photothermal efficiency. In addition, our approach can identify the contribution of the wavelength range of the incident light to the heating process. From this, we found that the temperature change caused by the energy of UV range is relatively small to damage the metamaterials. The degradation of metamaterials is only induced by UV-induced bond breaking. Our theoretical approach can be exploited to investigate the photothermal heating of plasmonic nanoparticles randomly dispersed in water or other materials, which have a dielectric function and absorption coefficient strongly sensitive to frequency.

There are some ideas to develop this model. First, our approach and many simulations \cite{3,4,5} assume that physical quantities such as specific heat capacity and thermal conductivity do not depend on temperature. This assumption may not be true when the systems, particularly polymers and polymer composites, are investigated over a wide temperature range \cite{38,39,40}. This is a main reason why theoretical predictions at low-laser intensity are much closer to experimental data than that at high-laser intensity. Taking into account the thermal dependence of physical quantities in our model and simulations is interesting but challenging to solve. Second, our theoretical model can be used to determine the thermal conductivity of materials by adjusting the parameter to obtain the best fit between theoretical and experimental $\Delta T$. This problem is promising and under study. Third, we completely ignore the thermal dissipation from the absorber to the external environment in this work. However, different conditions of heat transfer can cause nontrivial variations in the thermal distribution. 
\begin{acknowledgments}
This research is funded by Vietnam Academy of Science and Technology under grant number KHCBVL.05/22-23. 
\end{acknowledgments}
\section*{Conflicts of interest}
There are no conflicts to declare.

\end{document}